\documentclass[prl,superscriptaddress,showpacs,twocolumn]{revtex4}

\usepackage{graphicx}
\begin{document}

\title{Role of the $d-f$ Coulomb interaction in intermediate valence and Kondo systems:
a numerical renormalization group study}
\author{A. K. Zhuravlev$^a$, V. Yu. Irkhin$^{a,*}$ and M. I. Katsnelson$^b$ \\
%EndAName
$^a$Institute of Metal Physics, Ekaterinburg 620219, Russia \\
$^b$Institute for Molecules and Materials, Radboud University
Nijmegen, NL 6525 ED Nijmegen, The Netherlands}

%\maketitle

\begin{abstract}
Using numerical renormalization group method the temperature
dependences of the magnetic susceptibility $\chi(T)$ and specific
heat $C(T)$ are obtained in the single-impurity Anderson model
with inclusion of the $d-f$ Coulomb interaction. It is shown that
the exciton effects owing to this interaction can change
considerably the dependence $C(T)$ in comparison with the standard
Anderson model at not too low temperatures, whereas the dependence
$\chi (T)$ remains universal. The renormalization of the effective
hybridization parameter and $f$-level position, which is connected
with the $d-f$ interaction, is calculated, a satisfactory
agreement with the Hartree-Fock approximation being derived.
\end{abstract}

\maketitle

There is an interesting class of the $4f$-electron compounds
demonstrating intermediate valence (IV) of rare earth elements
(usually between 2+ and 3+) in a number of properties, e.g., in
the lattice constants (which are intermediate between those for
isostructural compounds with di- and trivalent ions), core-level
spectra (which are mixtures of the spectra for di- and trivalent
ions with comparable weights), and many others
\cite{conf1,conf2,rise1,rise2}. Heavy fermion (HF) compounds
\cite{stewart} form another important class of the $f$-electron
systems with anomalous properties. For the HF metals it is
commonly accepted now that they are the Kondo lattices, which
means that the small energy scale in the electron properties is
the Kondo temperature $T_K$,  i.e. the width of the Kondo
resonance owing to spin-dependent scattering of conduction
electrons by $f$-electron centers \cite{hewson}. As for the IV
compounds, they are frequently considered also as the Kondo
lattices, but just with higher $T_K$ (see, e.g.,
Ref.\onlinecite{rise2}).

Actually, such a consideration is not quite accurate since,
besides the spin (``Kondo'') fluctuations, valence or
\textit{charge} fluctuations should be also treated in such
systems. They are determined in part by the Coulomb repulson $G$
between conduction and localized electrons (the Falicov-Kimball
interaction \cite{falicov}). Taking into account the $d-f$
interaction together with the hybridization processes it is
possible to describe the IV state as a kind of exciton
condensation \cite{stevens,IK86}. Recently, the method of
first-principle calculations of the  parameter $G$ has been
proposed, and it was demonstrated that an account of this
interaction is necessary to describe properly the equation of
state for IV phase of Yb under pressure \cite{prlYb}.

At present, the usual Kondo effect is theoretically studied
thoroughly within the $s-d$ exchange (Kondo) and Anderson models.
Moreover, in the one-impurity situation the exact numerical
(renormalization group) \cite{Wilson,krishna} and analytical
Bethe-ansatz \cite{Wieg,Andrei} solution of this problem is obtained.
Universal curves describing the behavior of thermodynamic
properties were obtained for the Kondo \cite{Wilson,krishna} and
intermediate valence \cite{krishna} regimes, which permit a
detailed comparison with experimental data on anomalous
$f$-systems. At the same time, in the presence of both the $s-d$
exchange and  Coulomb interaction such a detailed information is
absent.

Formally, the charge fluctuations can be also described in terms
of a pseudo-Kondo effect, the states with (without) $f$-hole being
considered as pseudospin-up (down) states, respectively
\cite{WiegFink,schlottm}. It is the degeneracy of quantum states
for a scattering center which is important for the formation of
the Kondo resonance \cite{cox}. In the IV case the divalent and
trivalent states are degenerate by definition, thus this analogy
is not surprising. Therefore it is natural to consider the Kondo
phenomenon for the IV compounds taking into account both spin and
charge fluctuations, or, equivalently, both the ``Kondo'' and
exciton (``Falicov-Kimball'') effects. Since there is no clear
demarcation between the IV and Kondo systems, it can be supposed
that the exciton effects are relevant also for the latter case.
Recent analysis of the interplay of the true Kondo  and
pseudo-Kondo (exciton) effects \cite{prz} by the ``poor-man
scaling'' approach \cite{And,haldane} has demonstrated an
essential modification of the low-energy (infrared) behavior in
comparison with pure cases of the Anderson model and
Falicov-Kimball (``resonant level'') models. However, this
approach can give only a qualitative insight in the properties of
the system. Here we investigate the effects of this interplay by
applying  numerical renormalization group (NRG) approach
\cite{Wilson,krishna}.

We proceed with the Hamiltonian of the asymmetric Anderson model
with inclusion of the Falicov-Kimball interaction (on-site $d-f$
Coulomb repulsion $G$),
\begin{eqnarray}
\mathcal{H} &=&\sum_{\mathbf{k}\sigma }t_{\mathbf{k}}c_{\mathbf{k}\sigma
}^{\dagger }c_{\mathbf{k}\sigma }+\sum_\sigma [E_ff_\sigma ^{\dagger
}f_\sigma \ +V\left( c_\sigma ^{\dagger }f_\sigma +f_\sigma ^{\dagger }c_{%
\sigma }\right) ]  \nonumber \\
&&\ \ +G\sum_{\sigma \sigma ^{\prime }}f_\sigma ^{\dagger }f_\sigma
c_{\sigma ^{\prime }}^{\dagger }c_{\sigma ^{\prime }}
\end{eqnarray}
where the on-site $f-f$ Coulomb interaction $U$ is put to
infinity, so that the doubly occupied states are forbidden;
$f_{i\sigma }^{\dagger }=|i\sigma \rangle \langle i0|$ are the
Hubbard operator ($|i\sigma \rangle $ and $|i0\rangle $ are
single-occupied and empty site states); we neglect  for
simplicity $\mathbf{k}$-dependence of the hybridization matrix
element $V$.

We use the standard NRG method for the Anderson model
\cite{krishna} with some important modifications. At each NRG step
one obtains a finite-resolution spectrum which is truncated owing
to neglect of high-energy states \cite{Wilson}. Thus we have a
sequence of truncated energy spectra, the level resolution
decreasing with increasing iteration step. An automatic choose of
an optimal temperature for each NRG step is a main distinctive
feature of our calculations. Indeed, we cannot perform
calculations of thermodynamic averages at too low temperatures
since the discreteness of energy level leads to an uncontrollable
error. On the other hand, at sufficiently high temperatures the
high-energy states neglected can give an appreciable contribution
to the partition function, which is proportional to the factor of
$\exp\left( -E/k_BT \right)$. Therefore we estimate the
contribution of the upper states and increase the temperature to
make their contribution to be equal to the error chosen.

\begin{figure}[htbp]
\includegraphics[width=3.3in]{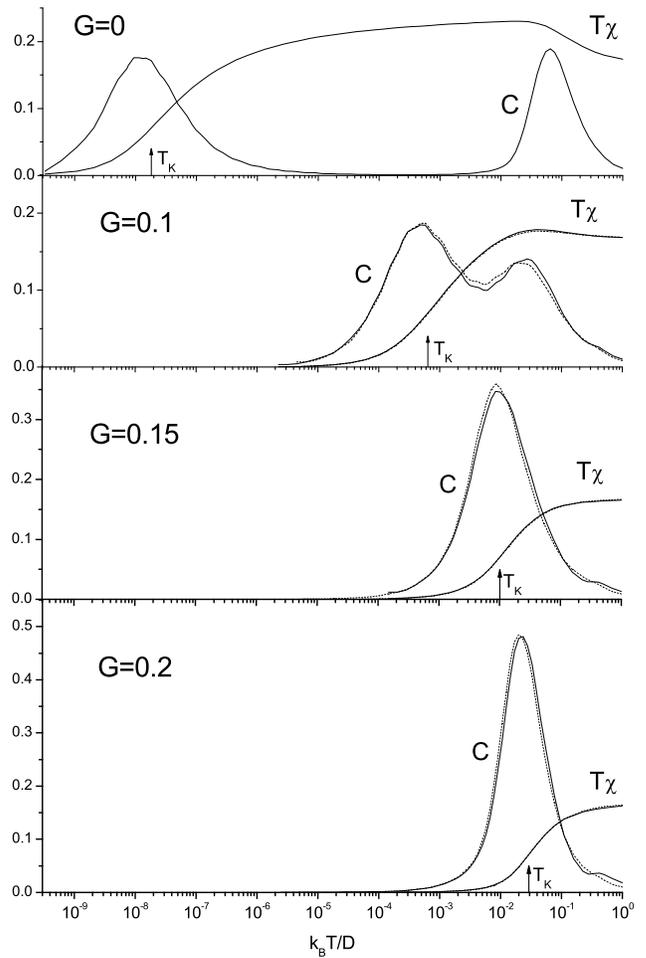}
\caption{Effective Curie constant $k_BT\chi(T)/(g\mu_B)^2$ and
specific heat $C(T)/k_B$ for $E_f=-0.14, V=0.1$. Solid line
corresponds to finite $G$, and the dotted line to the case $G$=0
with parameters $E_f^{(G=0)}$ and $V^{(G=0)}$ (see their values in
Table \ref{Tabl}). Below $T_K$ we have the universal Wilson
curve.} \label{FigE014}
\end{figure}

\begin{figure}[htbp]
\includegraphics[width=3.3in]{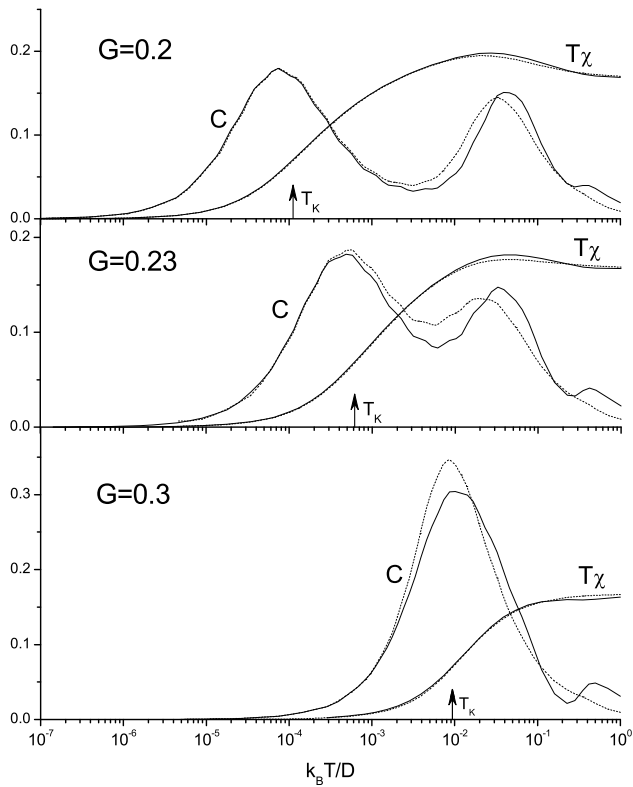}
\caption{The same data as in Fig. \ref{FigE014} for $E_f=-0.25,
V=0.1$.} \label{FigE025}
\end{figure}

We have used a rectangular conduction electron density of states
with the half-bandwidth of $D=1$. The magnetic susceptibility,
specific heat, and impurity level occupation number $n_f$
(valence) were calculated. Computational results are shown in the
figures. Figs. \ref{FigE014} and \ref{FigE025} demonstrate a
crossover from a two-maximum to one-maximum temperature behavior
of specific heat. It should be noted that such a crossover takes
place also in the standard Anderson model with changing $E_f$
\cite{Okiji}.

\begin{figure}[htbp]
\includegraphics[width=3.3in]{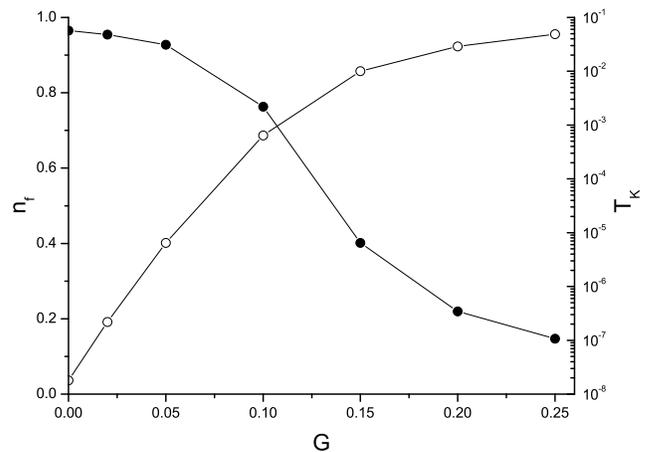}
\caption{The dependences of $n_f(\bullet)$ and $T_K(\circ)$ on $G$. ($E_f=-0.14,
V=0.1$). } \label{FigTKNf}
\end{figure}

\begin{table}[htbp]
\caption{The dependences of the Kondo temperature $T_K$ and
impurity occupation number $n_f$ on $G$ for $V=0.1$; the
Hartree-Fock values $V^{HF}$ and $E^{HF}$ are defined by Eq.
(\ref{HF}). The quantities $E_f^{(G=0)}$ and $V^{(G=0)}$ are
discussed in the text.}
\begin{tabular}{|c|c|c|c|c|c|c|c|}
\hline
$E_f$& $G$ &      $k_BT_K$      & $n_f$ & $E_f^{(G=0)}$&$E^{HF}_{f}$ & $V^{(G=0)}$ & $V^{HF}$\\ \hline
-0.06& 0   & 7.430$\cdot10^{-5}$&0.875&  -0.06     &-0.06 &0.1 &0.1\\
-0.06& 0.01& 2.290$\cdot10^{-4}$&0.825&  -0.049    &-0.050& 0.1  &0.1010            \\
-0.06& 0.02& 6.592$\cdot10^{-4}$&0.750&  -0.04     &-0.040& 0.1015&0.1024            \\
-0.06& 0.03& 1.703$\cdot10^{-3}$&0.651&  -0.031    &-0.0316& 0.103 &0.1041           \\
-0.06& 0.04& 3.714$\cdot10^{-3}$&0.542&  -0.022    &-0.021& 0.104&0.1059            \\
-0.06& 0.05& 6.798$\cdot10^{-3}$&0.443&  -0.013    &-0.011& 0.105&0.1078           \\ \hline

-0.14& 0   &1.813$\cdot10^{-8}$  &0.965  & -0.14     &-0.14   &0.1   &0.1\\
-0.14& 0.02&2.195$\cdot10^{-7}$&0.955& -0.120     &-0.1207&0.101 &0.1013            \\
-0.14& 0.05&6.452$\cdot10^{-6}$&0.928& -0.092    &-0.0934&0.105 &0.1039                \\
-0.14& 0.1 &6.416$\cdot10^{-4}$&0.763& -0.048    &-0.0500&0.11  & 0.1126             \\
-0.14& 0.15&1.003$\cdot10^{-2}$&0.402& -0.01    & 0.0008&0.116 &0.1250              \\
-0.14& 0.2 &2.882$\cdot10^{-2}$&0.220&  0.027   & 0.0535&0.12  &0.1316 \\  \hline

-0.25& 0.2 &1.115$\cdot10^{-4}$&0.865&-0.07      &-0.0947&0.111  &0.1219             \\
-0.25& 0.23&6.430$\cdot10^{-4}$&0.771&-0.048     &-0.0706&0.11  &0.1311             \\
-0.25& 0.3 &9.369$\cdot10^{-3}$&0.443&-0.013     & 0.008  &0.12   &0.1528             \\ \hline
\end{tabular}
\label{Tabl}
\end{table}

\begin{figure}[htbp]
\includegraphics[width=3.3in, angle=0]{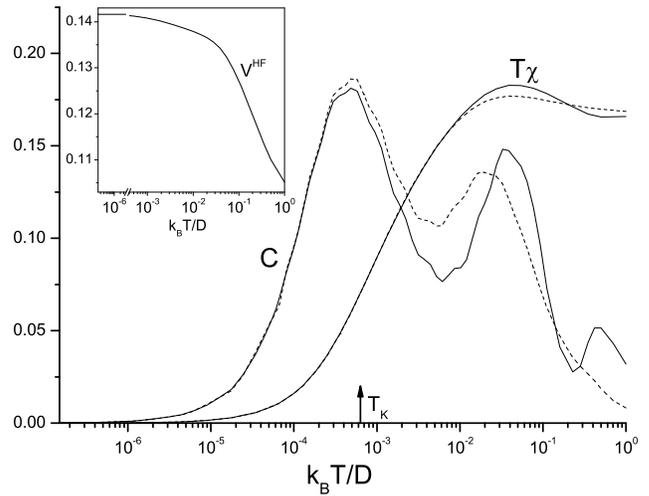}
\caption{Effective Curie constant $k_BT\chi(T)/(g\mu_B)^2$ and
specific heat $C(T)/k_B$ for $E_f$=-0.3, $V$=0.1, $G$=0.3 (solid)
and $E_f^{(G=0)}$=-0.0475, $V^{(G=0)}$=0.109, $G$=0 (dotted). For
these parameter sets we have $n_f$=0.77. Insert shows the
dependence $V^{HF}(T)$ according to Eq.(\ref{HF}).} \label{FigE03}
\end{figure}

\begin{figure}[htbp]
\includegraphics[width=3.3in, angle=0]{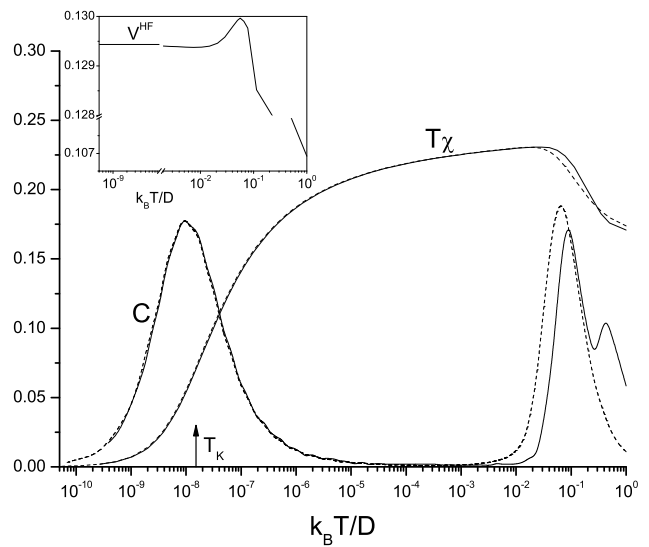}
\caption{Effective Curie constant $k_BT\chi(T)/(g\mu_B)^2$ and
specific heat $C(T)/k_B$ for $E_f$=-0.5, $V$=0.1, $G$=0.4 (solid)
and $E_f^{(G=0)}$=-0.14, $V^{(G=0)}$=0.1, $G$=0 (dotted). For
these parameter sets we have $n_f$=0.96 at $T$=0. Insert shows the
dependence $V^{HF}(T)$.} \label{FigE05}
\end{figure}

One can see that the temperature dependence of the magnetic
susceptibility is always similar to that in the Anderson model
without $d-f$ Coulomb interaction. At the same, the specific heat
behavior can be considerably different, especially for
sufficiently large $G$.

Similar to Ref.\onlinecite{krishna}, the Kondo temperature $T_K$
was determined from the condition $k_BT\chi(T)/(g\mu_B)^2 = 0.0701$ at
$T=T_K$. The dependence of the ground-state occupation number
of $f$-level $n_f$ and the Kondo temperature on $G$ are illustrated by Fig.
\ref{FigTKNf}. A more detailed information is
presented in the Table \ref{Tabl}.

There is an important question whether the effects of the $d-f$
Coulomb interaction can be described just by the renormalization
of the parameters usual Anderson Hamiltonian (without the Falicov
interaction)  or they can result in qualitatively new effects. To
investigate this problem we defined the effective hybridization
parameter $V^{(G=0)}$ and the effective position of the $f$-level,
$E^{(G=0)}_{f}$, as the parameters of the standard Anderson model
(with $G=0$) that gives the same values of $n_f$ (at zero
temperature) and the Kondo temperature, as our Hamiltonian (1). A
comparison of our computational results with those for the model
with $G=0$ and with the effective parameters introduced above
shows (Fig. \ref{FigE03} and \ref{FigE05}) that for the
susceptibility the effects of $G$ in the temperature dependence
can be completely eliminated by the parameter renormalization. At
the same time, for the specific heat this is, generally speaking,
possible for low enough temperatures, of order of $T_K$ or below.
This means that the Wilson ratio is not influenced by the $d-f$
interaction at $T\leq T_K$, but its ``temperature dependence'' at
higher temperatures is different for the cases $G=0$ and $G\neq
0$. Of course, it is not surprising that the $d-f$ Coulomb (but
not exchange) interaction is less important for the magnetic
susceptibility (which is connected only with spin degrees of
freedom) than for the specific heat  (which characterizes both
spin and charge fluctuations).

It is interesting to compare our renormalized model parameters
with their values from the unrestricted Hartree-Fock approximation \cite{IK86},

\begin{equation}
E^{HF}_{f} = E_f +G \sum_{\sigma }\langle c^{\dagger}_{\sigma}c_{\sigma}\rangle,
V^{HF} = V - G \langle c^{\dagger}_{\sigma}f_{\sigma}\rangle.
\label{HF}
\end{equation}
A comparison of the parameters of the effective Anderson model and
of the Hartree-Fock values (2), which are presented in the Table
\ref{Tabl}, shows that this approximation works well enough, at
least for not too large $d-f$ interaction ($G<0.25$). Thus
corresponding Coulomb correlation effects are not important. This
justifies the implementation of the Hartree-Fock approximation
into the first-principle electronic structure calculations in
Ref.\onlinecite{prlYb}. The dependences $V^{HF}(T)$ according to
Eq.(\ref{HF}) are shown in inserts in Figs.(\ref{FigE03}),
(\ref{FigE05}). One can see that in the Kondo regime a maximum
occurs which is qualitatively similar to the result of the
poor-man scaling consideration \cite{prz}.

To conclude, we have obtained an accurate NRG solution of the
one-impurity Anderson model with inclusion of the Falicov-Kimball
interaction (excitonic effects). Some  new features in comparison
with the standard Anderson model (in particular, in the
temperature dependence of specific heat) occur. A generalization
of the results to a lattice case would be of interest for the
theory of the Kondo lattices and IV compounds.

The research described was supported by Grant No. 747.2003.2 from
the Russian Basic Research Foundation (Support of Scientific
Schools), by Physical Division of RAS (agreement N
10104-71/OFN-03/032-348/140705-126), Ural Division of RAS (grant
for young scientists), and by the Netherlands Organization for
Scientific Research (Grant NWO 047.016.005).

\end{document}